\documentclass[8.5pt,twoside,twocolumn]{article}
\usepackage[super,sort&compress,comma]{natbib} %original Nanoscale 
\usepackage{mhchem}
\usepackage{times,mathptmx}
\usepackage{sectsty}
\usepackage{balance} 

\usepackage{graphicx} %eps figures can be used instead
\usepackage{lastpage}
\usepackage[format=plain,justification=raggedright,singlelinecheck=false,font=small,labelfont=bf,labelsep=space]{caption} 
\usepackage{fancyhdr}
\begin{document}

\thispagestyle{plain}
\fancypagestyle{plain}{
\fancyhead[L]{\includegraphics[height=8pt]{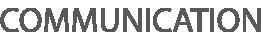}}
\fancyhead[C]{\hspace{-1cm}\includegraphics[height=20pt]{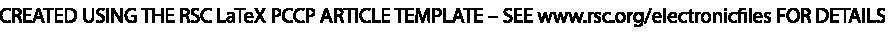}}
\fancyhead[R]{\includegraphics[height=10pt]{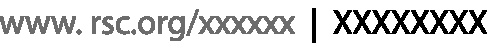}\vspace{-0.2cm}}
\renewcommand{\headrulewidth}{1pt}}
\renewcommand{\thefootnote}{\fnsymbol{footnote}}
\renewcommand\footnoterule{\vspace*{1pt}
\hrule width 3.4in height 0.4pt \vspace*{5pt}} 
\setcounter{secnumdepth}{5}

\makeatletter 
\def\subsubsection{\@startsection{subsubsection}{3}{10pt}{-1.25ex plus -1ex minus -.1ex}{0ex plus 0ex}{\normalsize\bf}} 
\def\paragraph{\@startsection{paragraph}{4}{10pt}{-1.25ex plus -1ex minus -.1ex}{0ex plus 0ex}{\normalsize\textit}} 
\renewcommand\@biblabel[1]{#1}            
\renewcommand\@makefntext[1]% 
{\noindent\makebox[0pt][r]{\@thefnmark\,}#1}
\makeatother 
\renewcommand{\figurename}{\small{Fig.}~}
\sectionfont{\large}
\subsectionfont{\normalsize} 

\fancyfoot{}
\fancyfoot[LO,RE]{\vspace{-7pt}\includegraphics[height=9pt]{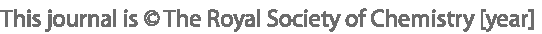}}
\fancyfoot[CO]{\vspace{-7.2pt}\hspace{12.2cm}\includegraphics{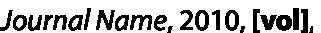}}
\fancyfoot[CE]{\vspace{-7.5pt}\hspace{-13.5cm}\includegraphics{RF}}
\fancyfoot[RO]{\footnotesize{\sffamily{1--\pageref{LastPage} ~\textbar  \hspace{2pt}\thepage}}}
\fancyfoot[LE]{\footnotesize{\sffamily{\thepage~\textbar\hspace{3.45cm} 1--\pageref{LastPage}}}}
\fancyhead{}
\renewcommand{\headrulewidth}{1pt} 
\renewcommand{\footrulewidth}{1pt}
\setlength{\arrayrulewidth}{1pt}
\setlength{\columnsep}{6.5mm}
\setlength\bibsep{1pt}

\twocolumn[
  \begin{@twocolumnfalse}
\noindent\LARGE{\textbf{STM-induced surface aggregates on metals and oxidized silicon}}%$^\dag$}}
\vspace{0.6cm}

\noindent\large{\textbf{Dominik St\"{o}ffler,$^{\ast}$\textit{$^{a}$} Hilbert von L\"ohneysen,\textit{$^{ab}$} and
Regina Hoffmann\textit{$^{a}$}}}\vspace{0.5cm}
%Please note that \ast indicates the corresponding author(s) but no footnote text is required. 

\noindent\textit{\small{\textbf{Received Xth XXXXXXXXXX 20XX, Accepted Xth XXXXXXXXX 20XX\newline
First published on the web Xth XXXXXXXXXX 200X}}}

\noindent \textbf{\small{DOI: 10.1039/b000000x}}
\vspace{0.6cm}
%Please do not change this text.

\noindent \normalsize{We have observed an aggregation of carbon or carbon derivatives on platinum and natively oxidized silicon surfaces during STM measurements in ultra-high vacuum on solvent-cleaned samples previously structured by e-beam lithography. We have imaged the aggregated layer with scanning tunneling microscopy (STM) as well as scanning electron microscopy (SEM). The amount of the aggregated material increases with the number of STM scans and with the tunneling voltage. Film thicknesses of up to 10 nm with five successive STM measurements have been obtained.}
\vspace{0.5cm}
 \end{@twocolumnfalse}
 ]

\section{Introduction}

With the ongoing process of miniaturization of electronic circuits, there is a growing interest to analyze quantum effects which play an important role in metal contacts of nanometer size. A common technique to pattern structures with sizes from the micrometer range down to tens of nanometers is e-beam lithography (EBL). The e-beam of a scanning electron microscope (SEM) is used to pattern an organic resist layer. Then a metal film is deposited onto the structure. Finally a lift-off process removes the remaining resist together with the unused part of the metallic layer such that metallic structures are left on the sample surface. Due to the small size of the structures, a characterization by scanning probe techniques is desirable. Samples studied by surface science methods such as scanning probe techniques are however often cleaned extensively under ultra-high vacuum conditions to avoid artifacts caused by contamination of the sample surface. On the other hand, in electronic transport studies of single organic molecules, solvent-cleaned EBL-fabricated metallic nanostructures for electrodes are often used under ambient pressure. It is commonly known that during SEM measurements carbon is often aggregated on the sample surface because of residual hydrocarbon molecules \cite{Miura98,Miura97}. This effect can even be exploited in ``e-beam contamination lithography'' \cite{Koops88}. Layers can also be aggregated using scanning tunneling microscopy (STM) to dissociate precursor gases \cite{Rauscher97}. It has also been reported that the electron current in STM measurements can dissociate oil vapor and deposit carbon on gold films under high-vacuum conditions \cite{Baba90}. For solvent-cleaned samples structured by EBL, we have observed the aggregation of a possibly carbon-rich layer on platinum, gold, and natively oxidized silicon surfaces during STM measurements in ultra-high vacuum, likely due to dissociation of molecular remnants from the EBL and lift-off processing. This so-called contamination resist has first been reported bei Ehrichs et al. \cite{Ehrichs88} when investigating the dissociation of organometallic gases in high vacuum (background pressure$<1 \times 10^{-8}$ mBar)  using a STM pulse technique.  Here we report on a systematic investigation of the aggregated layer with STM as well as SEM. The amount of the aggregated material increases with the number of scans and the tunneling voltage. Layer thicknesses of up to 10 nm with five successive STM measurements have been observed.
%Footnotes
%\footnotetext{\dag~Electronic Supplementary Information (ESI) available: [details of any supplementary information available should be included here]. See DOI: 10.1039/b000000x/}

%Please use \dag to cite the ESI in the main text of the article.
%If you article does not have ESI please remove the the \dag symbol from the title and the above footnotetext.

\footnotetext{\textit{$^{a}$Karlsruhe Institute of Technology - Campus South, Physikalisches Institut, D-76128 Karlsruhe, Germany Fax: +49 721 608 46103; Tel: +49 721 608 43520; E-mail: dominik.stoeffler@kit.edu \\
Karlsruhe Institute of Technology - Campus South, DFG-Center for Functional Nanostructures (CFN), D-76128 Karlsruhe, Germany}}
\footnotetext{\textit{$^{b}$Karlsruhe Institute of Technology - Campus North, Institut f\"ur Festk\"orperphysik, D76021 Karlsruhe, Germany}}

\section{Experimental methods}

\begin{figure}
\begin{center}
\includegraphics[width=7cm,angle=0,clip]{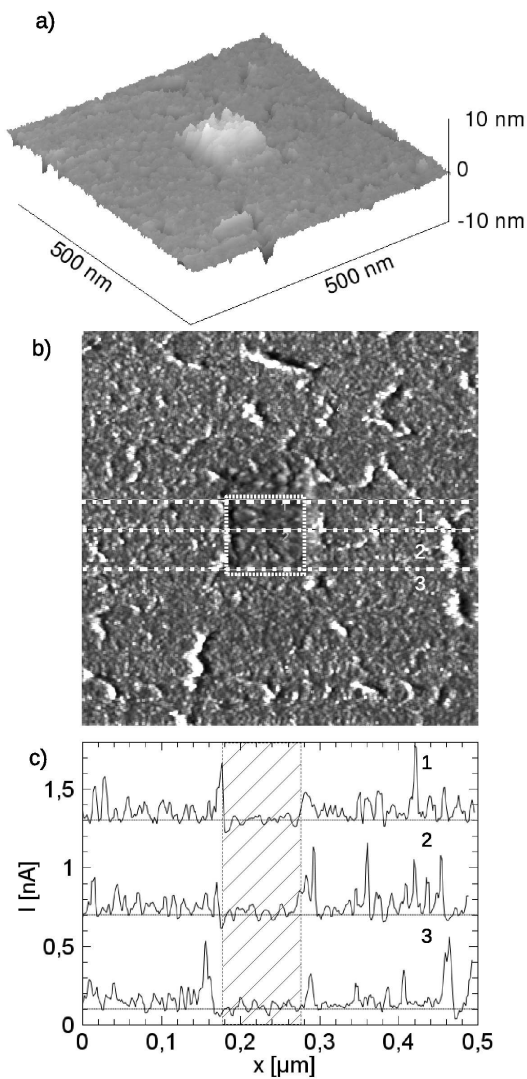}	
\end{center}
\vspace{-0.5cm}
\caption{a) STM image of a Pt film on Si, prepared as described in the main text ($U_T=8$ V, $I_T=0.1$ nA, 500$\times$500 nm$^2$). Prior to that scan the square-shaped hillock in the middle of the image was scanned five times with $U_T=10$ V, $I_T=0.5$ nA and a frame size of 100$\times$100 nm$^2$. b) Current image of the same 500$\times$500 nm$^2$ scan  (averaged over an area of 5 $\times$ 5 nm$^2$ with a median filter). The hillock region was observed in the area marked by a white square. The bright bands correspond to grain boundaries in the metallic film. c) Three examples of linescans taken along the white horizontal lines in the current image (shift of 0.6 nA between each profile). The deviations of the current from the setpoint (0.1 nA, horizontal lines in profiles) are significantly lower in the hillock region (hatched).}
\label{stoe2_f1}
\end{figure}

We used Si(001) substrates with a native oxide layer. We then deposited a PMMA resist layer on the sample surface. To create the pattern of metal wires and contact pads on the surface we first wrote structures of lateral dimensions between $\sim$ 50 nm and 100 $\mu$ m with a commercial Zeiss SEM using a Raith e-beam lithography system. We then treated the sample with methyl isobutyl ketone (MIBK) as developer and with isopropanol, and electron-beam deposited a 20-nm thick metallic layer (Au or Pt) in a high-vacuum evaporation chamber (background pressure $<1 \times 10^{-8}$ mBar) on the patterned surface. For lift-off we inserted the sample into fresh acetone heated to near its boiling point for approximately 5 min. We then repeatedly rinsed the sample using room-temperature acetone. Between e-beam writing, evaporation, and lift-off the sample was exposed to air for several days. In order to contact the metallic structures on the sample for STM imaging, Au wires were attached with conductive, UHV-compatible glue (EpoTek) to the sample surface. These contacts were located at a distance of about 1 mm from the area scanned in STM images. The sample was then transferred to an Omicron LT-STM with a background pressure smaller than $5 \times 10^{-10}$ mbar without further treatment in vacuum. We prepared tungsten tips by sputtering and electron-beam heating in situ. We then approached the freshly prepared tip to the metallized part of the surface. We used a commercial  Nanonis electronics and a Femto low-noise preamplifier for STM imaging at room temperature.

\vspace{-0.5 cm}
\begin{figure}
        \begin{center}
	    \includegraphics[width=\linewidth,angle=0,clip]{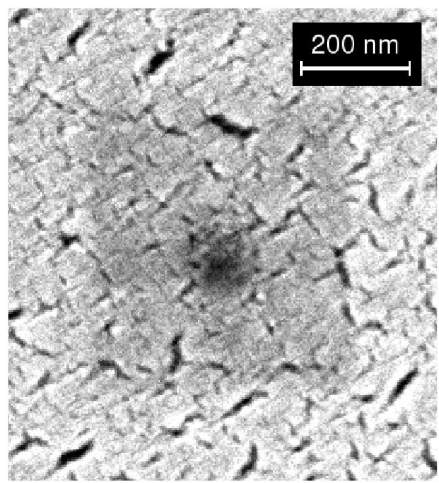}
        \end{center}
\vspace{-0.5cm}
\caption{SEM image of the same sample we had previously scanned with STM (see image in Fig. \ref{stoe2_f1} b). The dark regions match the STM frame sizes with an error below 6 $\%$. The structures in the metal film are due to the polycrystalline growth of Pt on SiO$_x$. Dark features in the SEM image correspond to bright features in the STM current image (Fig. \ref{stoe2_f1} b).}
	\label{stoe2_f3}
\end{figure}

\section{Results and discussion}

In spite of the insulating oxide layer of a few nanometer thickness \cite{Morita90} on the silicon substrate it was possible to perform STM measurements on the sample, even on areas not covered by metal. On the metallic regions we used tunneling voltages between 4 V and 10 V, tunneling currents $I_T$ between 0.1 nA and 0.8 nA and scan speeds of up to 500 nm/s. Only positive sample voltages allowed a stable scanning. On the silicon oxide, voltages $\geq$ 8~V were necessary to tunnel through the oxide layer and thus to avoid a possible hard contact of the tip with the surface. Short current peaks of up to 10 nA occurred repeatedly during the measurement, which we relate to the presence of mobile molecular remnants on the surface, which are prone to being picked up by the STM tip during the measurement. Although we had thoroughly cleaned the sample with acetone after the fabrication process it is likely that molecular remnants were still present on the sample surface. When we used values of $I_T$ larger than 0.8 nA, we were not able to stabilize the tip position with our standard current feedback loop which varies the tip-sample distance to maintain a constant tunneling current. After taking a few STM images of the same area we observe the formation of hillocks. 

To investigate the aggregated material and the aggregation process by SEM we acquired several STM images at particular locations with respect to the pattern of metal structures made by e-beam lithography on the surface. We first scanned one particular area of 100 $\times$ 100 nm$^2$ five times using $U_T=10$~V, $I_T=0.5$~nA. We then performed a larger scan of 500 $\times$ 500 nm$^2$ with $U_T=8$~V, $I_T=0.1$~nA, which included the smaller area scanned before. The resulting image showed a hillock with a maximum height of 10 nm (Fig. \ref{stoe2_f1} a). This hillock has an enhanced surface roughness of about 1.5 nm peak-to-peak compared to the roughness of the evaporated metal surface which is about 0.5 nm peak-to-peak. 
We observe large current peaks of up to 10 nA, i.e., the range limit of the preamplifier, during the 2 ms recording of individual pixels of 1 nm. We averaged these current peaks with a median filter, and show the resulting current image in Fig. \ref{stoe2_f1} b. These averaged current peaks are visible in Fig. \ref{stoe2_f1} b as small white spots. The extended features in Fig. \ref{stoe2_f1} b are due to the polycrystalline granular structure of the Pt film (see also Fig \ref{stoe2_f3}). We observed a high density of current peaks all over the image, except for the region where the 100 $\times$ 100 nm$^2$ scans (dotted square in the image) had been obtained. There the current is stabilized better with less noise at the setpoint of 0.1 nA by the feedback loop. This can be seen more clearly in line profiles (Fig. \ref{stoe2_f1} c), taken along the dotted horizontal lines in Fig. \ref{stoe2_f1} b. A control experiment with the same parameters and scanning frame sizes on an evaporated gold film on silicon showed no aggregation and, furthermore, a much lower noise level with peaks below 1 nA in the tunneling current compared to the 10-nA peaks on the e-beam lithography prepared samples.

We then transferred the sample to the SEM on the same day as removing it from the LT-STM vacuum chamber. Due to the layout of the structures it was possible to identify in SEM images the region of the STM scans  and thus to relate the images obtained by both methods. The STM-scanned areas appeared as regions with a reduced brightness in the SEM images. The 500$\times$500 nm$^2$ STM scan was visible as a lightly darkened square on the platinum layer with an even darker square in the center, where the five consecutive 100 $\times$ 100 nm$^2$ scans had been performed (Fig. \ref{stoe2_f3}). The size of these dark regions matches the original STM scanning frame size. A variation of the acceleration voltage (1, 3 and 10 keV) does not show a corresponding systematic contrast change in our images.

We first discuss whether material transfered from the tip to the surface \cite{Mamin91} or vice versa could be the origin of the observed effects. We expect that the applied voltages and the currents we used for the feedback control are not sufficient for electromigration processes. If material transfer from the sample to the tip is the origin of the hillocks, indentations should be observed on the surface instead of protrusions. One might argue that often in STM images, due to electronic effects, topographically indented regions can appear as protrusions and vice versa. We can exclude this possibility because the aggregates were also observed on the parts of the sample covered by native SiO$_x$, with similar features as on the metal-covered regions.

In case of a possible material transport from the tip to the sample, tungsten from the tip should be found on the sample surface which can be detected by energy dispersive X-ray spectroscopy (EDX). In order to check this possibility, we fabricated an extended region covered by the aggregated material on the native silicon oxide, using a gap voltage of 10 V and a tunneling voltage of 0.5 nA to scan three frames with a size of 1500 $\times$ 1500 nm$^2$, laterally shifted by the piezo scanner. We detected in SEM images that material was indeed aggregated on the silicon oxide with a similar appearance as the other aggregated structures. This excludes the possibility of metal oxid creation and aggregation through water layer dissociation as well. EDX measurements on this area showed no traces of any metal (neither W from the tip, nor Au or Pt from the sample). We therefore conclude that nonmetallic remnants from the fabrication process are a likely origin of the aggregate. 

Binding energies of organic molecules are usually of the order of a few eV. It is therefore possible that with large tunneling voltages of several eV, electron transfer to or from the molecule, leading to dissociation, may occur. In order to investigate this possibility, we studied the dependence of the aggregation process on the tunneling parameters.

\begin{figure}
        \begin{center}
	    \includegraphics[width=\linewidth,angle=0,clip]{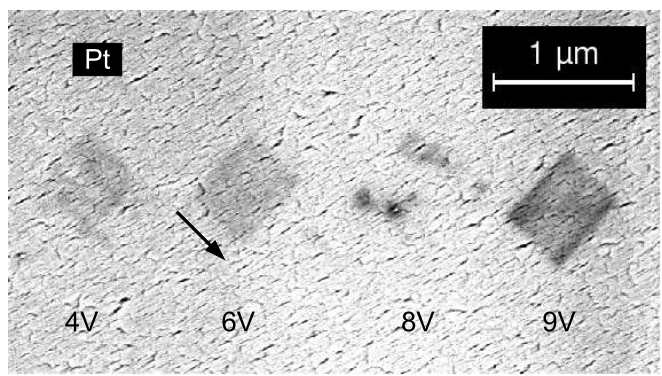}
        \end{center}
\vspace{-0.5cm}
\caption{SEM image at 10 keV. Four STM scanned regions with 500$\times$500 nm$^2$ at an angle of 45$^\circ$ can be seen as darker regions on the platinum surface ($I_T=0.2$ nA). The arrow indicates the STM line scan direction.}
	\label{stoe2_f2}
\end{figure}

Scans with a fixed $U_T=9$ V and $I_T=$ 0.1, 0.2, 0.4, 0.5 nA appeared similar to each other in the SEM images (not shown). We additionally acquired four STM scans with different voltages ($U_T= 4, \; 6, \; 8,\; 9$ V) at the same $I_T=0.2$ nA setpoint. In order to exclude artifacts, we chose an angle of 45$^{\circ}$ with respect to the SEM scan on the platinum region (Fig. \ref{stoe2_f2}). With increasing voltages, the observed contrast increases. For some voltages, however, the STM-induced structures are not homogeneously distributed in the scanned areas, which can be seen clearly in the 8-V image whereas the 9-V image appears rather homogeneous. We furthermore measured the map of the tunneling current, our feedback parameter, where we observed several current peaks of up to 10 nA during scanning due to instabilities in the tunneling current, similar to the current fluctuations, reported in Fig. \ref{stoe2_f1}b above.

While we can exclude metal transfer from the tip to the surface or vice versa it is conceivable that molecular remnants on the sample surface jump to the tip and back.
The tip is surrounded by a large inhomogeneous electric field which can attract or repel polar molecules. If molecules jump we expect additional high-frequency noise in the tunneling current \cite{Berner01}. This could explain the current peaks. An inhomogeneous distribution of molecular remnants could explain why the structures are not homogeneously distributed in the scanned areas.

We further checked this hypothesis by analyzing the tunneling current in the scan taken at $U_T=8$~V where strong inhomogeneities in the scanned window are observed. While in the brighter regions the tunneling current fluctuated between 0.2 and 3 nA, in the dark regions large peaks around 8-10~nA appear in $I_T$. For the $U_T=9$ V image the current peaks were more homogeneously distributed and thus a more homogeneous aggregation can be seen in Fig. \ref{stoe2_f2} as well, in line with our interpretation.
From the systematic current map of Fig. \ref{stoe2_f1} b and c we can exclude feedback-control problems as origin of the current peaks because they do not occur in the central region where several STM scans were performed as opposed to the other regions. We interpret this surprising finding as follows. In the region where the several consecutive 100 $\times$ 100 nm$^2$ scans were conducted and the hillock was observed, a large fraction of the molecules have already dissociated and the aggregated carbon rests immobilized on the surface. It is only in the surrounding region where a sizable fraction of molecules is left which can dissociate and thus create peaks in the tunneling current.

Scanning tunneling spectroscopy data showed the same behaviour on both the hillock and the surrounding one-scan region, which is a characteristic field emission measurement between a tungsten tip and a platinum surface (\cite{Young71}, \cite{Beebe06}). Because of the organic molecules which are left over from the fabrication process, and the similar appearance of the substance in SEM images to the usual carbon-rich aggregation in SEM imaging, we suppose the aggregated material to be carbon-rich. Unfortunately, EDX measurements were not able to detect a difference of carbon content between hillocks and the rest of the sample because of the EDX-induced carbon accumulation. It is well established that platinum binds with carbon \cite{Kiguchi08}, e.g., in catalytic processes. However, we have observed qualitatively the same STM-induced topography change on Au surfaces as well.

We exclude an aggregation from the residual gas in our case, because the background pressure of our chamber is very low. Additionally, we have performed STM and high-resolution scanning force microscopy measurements on similar samples prepared by evaporation through stencil masks \cite{Stoeffler10}. These measurements show no indication of the presence of molecules on the surface. It has been shown that a UHV-STM can induce modifications in thin carbon films which can be used to fabricate nanostructures \cite{Muehl04}. It is possible that mobile molecular remnants are accumulated by the electric field of the STM tip without dissociation. If this were the only effect, we would expect a different topology of the hillocks on the surface, where material should be aggregating preferentially at the edge of an STM image, perpendicular to the line scan direction, which is indicated as arrow in Fig. \ref{stoe2_f2}. On the other hand, a chemical reaction taking place in the vicinity of the tip would lead to a more homogeneous distribution of remnants over the STM-scanned region, as observed at least for some voltages. Our data suggest that by using a scanning tunneling microscope in UHV one can in principle fabricate carbon-rich nanostructures with heights of up to 10 nm not only from the gas phase but also from molecular remnants preaggregated on the sample surface. In most cases such an aggregation is, however, unwanted and can lead to artifacts in determining the topography of nanostructures created by e-beam lithography.

\section{Conclusions}

In conclusion, we showed that during STM measurements on samples produced by e-beam lithography a layer of possibly carbon-rich material is aggregated on the surface, due to organic residues on the sample even under ultra-high vacuum conditions. The aggregation process changes the topography and may affect the physical properties of the surfaces. The amount of aggregated material depends on the tunneling voltage. Large tunneling voltages and eventually residues on the surface lead to feedback stabilization difficulties and therefore to large tunneling current peaks. The thickness of the STM-induced layer grows with the number of scans and can be as thick as 10 nm on platinum. \\
\vspace{-0.5cm}
\section*{ACKNOWLEDGMENTS}

We thank Dr. Christoph S\"urgers for useful discussions, the Landesstiftung Baden-W\"urttemberg, the Centre for Functional Nanostructures (CFN) and the ERC-project "Nanocontacts" for financial support.

\footnotesize{
\bibliography{stoe_nanoscale} %your .bib file
\bibliographystyle{rsc} %the RSC's .bst file
}

\end{document}